\documentclass[%
reprint,
superscriptaddress,
showpacs,preprintnumbers,
 amsmath,amssymb,
aps,
pre,
floatfix%,
]{revtex4-1}
\usepackage{amsmath}
\usepackage{graphicx}
\DeclareMathSizes{12}{17.28}{9}{7}

\usepackage{float}
\usepackage{dcolumn}% Align table columns on decimal point
\usepackage{bm}% bold math
\usepackage{bbold}
\usepackage[multidot]{grffile}
\usepackage[colorlinks=true, breaklinks=true]{hyperref}% add hypertext capabilities
\usepackage[mathlines]{lineno}% Enable numbering of text and display math
\renewcommand{\min}{{\mathrm{min}}}
\renewcommand{\max}{{\mathrm{max}}}

\usepackage{xcolor}

\usepackage{xspace}

\usepackage[utf8]{inputenc}
\usepackage{tikz}
\usetikzlibrary{shapes.geometric,fit,arrows,babel}
\usepackage[utf8]{inputenc} % allow utf-8 input
\usepackage[T1]{fontenc}    % use 8-bit T1 fonts
\usepackage{hyperref}
\hypersetup{colorlinks,
citecolor=blue
}
\usepackage{url}            % simple URL typesetting
\usepackage{booktabs}       % professional-quality tables
\usepackage{amsfonts}       % blackboard math symbols
\usepackage{nicefrac}       % compact symbols for 1/2, etc.
\usepackage{microtype}      % microtypography
\usepackage{lipsum}
\usepackage{amsmath,amssymb,amsthm}
\usepackage[normalem]{ulem}
\usepackage[colorinlistoftodos,prependcaption,textsize=tiny]{todonotes}
\usepackage{graphicx}
\usepackage{wrapfig}

\definecolor{Agreen}{rgb}{0.1, 0.6, 0.1} % define new green
\begin{document}

%\preprint{APS/123-QED}

\title{Diversity-induced decoherence}

\author{Marius E. Yamakou}
\email{marius.yamakou@fau.de}
\affiliation{Department of Data Science, Friedrich-Alexander-Universit\"{a}t Erlangen-N\"{u}rnberg, Cauerstr. 11, 91058 Erlangen, Germany}

\author{Els Heinsalu}
\email{els.heinsalu@kbfi.ee}
\affiliation{National Institute of Chemical Physics and Biophysics - Akadeemia tee 23, 12618 Tallinn, Estonia}
\author{Marco Patriarca}
\email{marco.patriarca@kbfi.ee}
\affiliation{National Institute of Chemical Physics and Biophysics - Akadeemia tee 23, 12618 Tallinn, Estonia}

\author{Stefano Scialla}
\email{stefano.scialla@kbfi.ee}
\affiliation{National Institute of Chemical Physics and Biophysics - Akadeemia tee 23, 12618 Tallinn, Estonia}
\affiliation{Department of Engineering, Universit\`a Campus Bio-Medico di Roma - Via \'A. del Portillo 21, 00128 Rome, Italy}

\begin{abstract} 
We analyze the effect of small-amplitude noise and heterogeneity in a network of coupled excitable oscillators with strong time scale separation. 
Using mean-field analysis, we uncover the mechanism of a new nontrivial effect --- \textit{diversity-induced decoherence} (DIDC) ---  in which heterogeneity modulates the mechanism of self-induced stochastic resonance to inhibit the coherence of oscillations. 
We argue that DIDC may offer one possible mechanism via which, in excitable neural systems, generic heterogeneity and background noise can synergistically prevent unwanted resonances that may be related to hyperkinetic movement disorders.
\end{abstract}

%%%%%%%%%%%%%%%%%%%%%%%%%%%%%%%%%%%%%%%%%%%%%%%%%%%%%%%%%%%%%%%%%%%%%%%%%%%%%%%%%%%%%%%%%%%%%%%%%%%%
%\pacs{05.45.Xt, 05.65.+b, 89.75.Fb}% PACS, the Physics and Astronomy
\maketitle

%%%%%%%%%%%%%%%%%%%%%%%%%%%%%%%%%%%%%%%%%%%%%%%%%%%%%%%%%%%%%%%%%%%%%%%%%%%%%%%%%%%%%%%%%%%%%%%%%%%%
%\section{Introduction}\label{Sec. I}
%%%%%%%%%%%%%%%%%%%%%%%%%%%%%%%%%%%%%%%%%%%%%%%%%%%%%%%%%%%%%%%%%%%%%%%%%%%%%%%%%%%%%%%%%%%%%%%%%%%%
The role of disorder in the dynamics of complex networks has been extensively studied in terms of noise and diversity (i.e., heterogeneity) effects~\cite{jung1995spatiotemporal,liu1999stochastic,busch2003influence,gammaitoni1998stochastic, mcdonnell2009stochastic,perez2010constructive}. For example, Shibata and Kaneko showed that heterogeneity enhances regularity in the collective dynamics of coupled map lattices, even if each element has chaotic dynamics~\cite{shibata1997heterogeneity}.
Later on, Cartwright observed the emergence of collective
network oscillations in a cubic lattice of locally coupled and diverse
FitzHugh-Nagumo (FHN) units, none of which were individually in an
oscillatory state~\cite{cartwright2000emergent}.
Tessone et al. demonstrated an amplification of the response of a coupled oscillator network to an external signal, driven by an optimal level of heterogeneity of its elements, and named this effect diversity-induced resonance (DIR)~\cite{tessone2006diversity,toral2009diversity,chen2009threshold,wu2010diversity,wu2010effects,patriarca2012diversity,tessone2013diversity,grace2014pattern,patriarca2015constructive,liang2020diversity}.
Other authors showed that DIR can occur even in the absence of an external forcing~\cite{kamal2015dynamic,scialla2021hubs}. Some of these studies concluded that stochastic resonance (SR) and DIR are substantially analogous phenomena~\cite{tessone2006diversity,tessone2007theory}, to the point that diversity may be viewed as a form of  ``quenched noise''.

Diversity in complex networks dynamics has been studied also in terms of its interaction with noise, by introducing in a system both types of disorder. Most of this research highlighted the possibility to amplify resonance effects caused by noise thanks to diversity optimization, and vice versa~\cite{boschi2001triggering,li2012parameter,li2014multiple,gassel2007doubly}.  Recently, Scialla et al.~\cite{scialla2022interplay} showed that the impact of diversity on network dynamics can be significantly different from that of noise and may result in an antagonistic effect, depending on the specific network configuration.  
At the same time, however, various regions of synergy between the two types of disorder, giving rise to strong resonance effects, were observed. Also, it has been shown that diversity in a network of FHN neurons can enhance coherence resonance (CR) \cite{zhou2001array}, which is a regular response (i.e., a limit cycle behavior) to an optimal noise amplitude \cite{pikovsky1997coherence}, occurring when the system is bounded near the bifurcation thresholds \cite{neiman1997coherence,liu2010multiple}.

Another form of noise-induced resonance is self-induced stochastic resonance (SISR), which has a different mechanism from CR for the emergence of regular oscillations \cite{deville2005two,yamakou2019control}.
SISR occurs when a small-amplitude noise perturbing the fast variable of an excitable system with a strong time scale separation results in the onset of coherent oscillations \cite{muratov2005self,yamakou2019control}. Due to the peculiarity of operating at relatively weak noise, SISR represents a particularly interesting case to study the effects of the interplay between noise and diversity. This is relevant to the potential role of SISR as a signal amplification mechanism in biological systems, given that diversity is inherent to networks of neurons or other cells.

In this Letter, we demonstrate that in contrast to previous literature, showing that network diversity can be optimized to enhance collective behaviors such as synchronization or coherence~\cite{shibata1997heterogeneity,cartwright2000emergent,tessone2006diversity,toral2009diversity,chen2009threshold,wu2010diversity,wu2010effects,patriarca2012diversity,tessone2013diversity,grace2014pattern,patriarca2015constructive,liang2020diversity,kamal2015dynamic,scialla2021hubs,tessone2007theory,scialla2022interplay,zhou2001array}, the effect of diversity on SISR, instead, can only be antagonistic. This indicates that the enhancement or deterioration of a noise-induced resonance phenomenon by diversity strongly depends on the underlying mechanism.

We point out that not only constructive, but also destructive resonance effects may have significant biological consequences. 
For instance, an increasing number of studies on Parkinson's disease~\cite{vogt2016diversity} indicate that dopaminergic neurons are characterized by a relatively high degree of heterogeneity, and  disease progression is associated with the death of only one or a few specific dopaminergic neuron subpopulations, leading to a loss of neuron diversity with respect to healthy brain tissues. 
Thus,  the role of diversity in biological systems might be also to inhibit unwanted resonances through compensatory mechanisms between different neuron sub-types, which can result in pathological conditions, if missing.

There is still a very limited understanding of the named phenomena from a complex systems modeling viewpoint, as previous works have focused mostly on systems and conditions that favor constructive resonance effects. 
In this work, we uncover   \textit{diversity-induced decoherence} (DIDC) mechanism, where, in contrast to its effect on CR, diversity deteriorates the coherence of oscillations due to SISR.

%%%%%%%%%%%%%%%%%%%%%%%%%%%%%%%%%%%%%%%%%%%%%%%%%%%%%%%%%%%%%%%%%%%%%%%%%%%%%%%%%%%%%%%%%%%%%%%%%%%%
%\section{Network Model}\label{Sec. II}
%%%%%%%%%%%%%%%%%%%%%%%%%%%%%%%%%%%%%%%%%%%%%%%%%%%%%%%%%%%%%%%%%%%%%%%%%%%%%%%%%%%%%%%%%%%%%%%%%%%%
As a paradigmatic model with well-known biological relevance, we  study the effects of diversity in a network of globally coupled FHN units~\cite{fitzhugh1960thresholds,fitzhugh1961impulses,nagumo1962active}:
\begin{eqnarray}\label{eq:1}
%\begin{split}
\left\{\begin{array}{lcl}
\displaystyle{\frac{dv_i}{dt}}&=&v_i\big(a_i-v_i\big)\big(v_i-1\big)-w_i\\[3.0mm]
&+&\displaystyle{K\sum\limits_{j=1}^{N}\big(v_j-v_i\big) + \eta_i(t)},\\[1.0mm]
\displaystyle{\frac{dw_i}{dt}}&=&\varepsilon\big(bv_i-cw_i\big).
\end{array}\right.
%\end{split}
\end{eqnarray}
Here $(v_i,w_i)\in\mathbb{R}^2$ represent the fast membrane potential and slow recovery current variables of the elements, respectively; the index $i=1,...,N$ stands for nodes;
$K>0$ is the synaptic coupling strength; $0<\varepsilon\ll1$ is the time scale separation between $v_i$ and $w_i$ and $b, c > 0$ are constant parameters. 
Diversity is introduced by assigning to each network element $i$ a different value of $a_i$, as specified below. 
The terms $\eta_i$ ($i=1,...,N$) are independent Gaussian noises with zero
mean, standard deviation $\sigma_{n}$, and correlation function $\langle \eta_i(t),\eta_j(t') \rangle= \sigma_{n}^2\delta_{ij}(t-t')$.
The noise intensity applied to each neuron will be measured by $\sigma_{n}$.

%%%%%%%%%%%%%%%%%%%%%%%%%%%%%%%%%%%%%%%%%%%%%%%%%%%%%%%%%%%%
%\section{Analysis}\label{Sec. III}
%%%%%%%%%%%%%%%%%%%%%%%%%%%%%%%%%%%%%%%%%%%%%%%%%%%%%%%%%%%%
The excitable regime where the network defined by Eqs.~\eqref{eq:1} has a \textit{unique} and \textit{stable} fixed point is the required deterministic state for the occurrence of SISR \cite{deville2007self,yamakou2018coherent,yamakou2020optimal}.
When $\eta_i=0$, the point $(v,w)=(0,0)$ becomes a fixed point of Eqs.~\eqref{eq:1}, and is unique if and only if 
\begin{equation}\label{eq:2}
\frac{(a_i-1)^2}{4}<\frac{b}{c}.
\end{equation}
For the fixed point $(v_f,w_f)=(0,0)$ to be stable, we must have $\mathrm{tr}J_{ij}<0$ and $\mathrm{det} 
J_{ij}>0$, where $J_{ij}$ is the Jacobian matrix of the linearized Eqs.~(1). 
Since $\varepsilon,c >0$, we have $\mathrm{tr}J_{ij}<0$ and $\mathrm{det}J_{ij}>0$ only if 
\begin{equation}\label{eq:4}
-3v_f^2 +2(a_i+1)v_f -a_i<0.
\end{equation}
To ensure that the network defined by Eqs.~\eqref{eq:1} lies in the excitable regime required for SISR,
in the following we set $b=1$ and $c=2$.
We also set $\varepsilon=0.001$, $K=0.1$, and $N=100$. To introduce diversity, the values of $a_i$ are drawn from a truncated Gaussian distribution in the interval $a_i\in(0,1+\sqrt{2})$, and are randomly assigned to network elements. 
The standard deviation $\sigma_{d}$ and mean $a_{m}$ of the distribution  measure diversity and how far the network is from the oscillatory regime (corresponding to $a_i\leq0$), respectively.

To study the effects of diversity $\sigma_{d}$ on SISR analytically, we apply the mean field approach introducing the global variables  $V(t)=N^{-1}\sum_{i=1}^{N}v_i(t)$ and $W(t)=N^{-1}\sum_{i=1}^{N}w_i(t)$.  
Adapting the method used in Refs.~\cite{desai1978statistical,tessone2006diversity,scialla2022interplay}, 
we set $v_i = V + \delta_i$ in Eqs.~\eqref{eq:1}, alongside the assumptions that $\sum_{i=1}^{N}\delta_i\approxeq 0$, $\sum_{i=1}^{N}\delta_i^3\approxeq 0$. 

We further assume that the standard deviation $\sigma_d$ of the $a_i$ distribution is small, allowing the approximation
\begin{equation}\label{eq:approx}
\left \langle a_i[(V+\delta_i)^2-(V+\delta_i)] \right \rangle \approxeq \left \langle a_i \right \rangle \left \langle (V+\delta_i)^2-(V+\delta_i) \right \rangle,
\end{equation}
where $\langle \dots \rangle$ denotes an average over the $N$ neurons. 
We note that the Gaussian distribution of $a_i\sim \mathcal{N}(a_m,\,\sigma_d)$ in the range $(0,1+\sqrt{2})$ is always truncated whenever a given value of $a_m$ and/or $\sigma_d$ pushes $a_i$ out of bounds, especially when $a_m$ is very close to the boundaries of $(0,1+\sqrt{2})$.

Using these assumptions and averaging Eqs.~\eqref{eq:1} over the $N$ neurons, we obtain the following dynamical equations for the global variables $V$ and $W$: 
\begin{eqnarray}\label{eq:5}
%\begin{split}
    \left\{
    \begin{array}{lcl}
    \displaystyle{\frac{dV}{dt}}&=&V \big[ \big(A-V \big) \big( V-1 \big) - 3M \big]\\[1.0mm]
    &+& \displaystyle{M(A+1) - W + \eta_{_G}(t)},\\[1.0mm]
    \displaystyle{\frac{dW}{dt}} &=& \varepsilon \big( bV - cW \big),
    \end{array}
    \right.
%\end{split}
\end{eqnarray}
where $M=N^{-1}\sum_{i=1}^{N}\delta_i^2$ and $A=N^{-1}\sum_{i=1}^{N}a_i$. 
$M$ can be considered as a diversity parameter, in that it increases with diversity in the network and $M=0$ for a homogeneous system ($\sigma_d=0$).
Noise effects are represented by a global white noise term $\eta_{_{G}}=N^{-1}\sum_{i=1}^{N}\eta_i$ with zero mean and correlation
function  $\langle \eta_{_G}(t),\eta_{_G}(t') \rangle= N^{-1}\sigma_{n}^2\delta(t-t')$.

When there is no noise in first equation of Eqs.~\eqref{eq:5},
$\eta_{_G}(t)=0$, then in the adiabatic limit $\varepsilon\to0$, for any initial condition of Eqs.~\eqref{eq:5}  the system relaxes to $V=V^{*}_{R}(W)$ and then to $V=V^{*}_{L}(W)$, where $V^{*}_{R}(W)$ and $V^{*}_{L}(W)$ are the right and left stable branches of the $V$-nullcline, respectively.
Solving $V\big[\big(A-V\big)\big(V-1\big)-3M\big]+ M(A+1)-W=0$ for $V$, we get three real and ordered solutions, namely, $V^{*}_{L}(W) \leq V^{*}_{S}(W) \leq V^{*}_{R}(W)$, which are all functions of $W$.

\begin{figure}
    \includegraphics[width=9.0cm,height=5.0cm]{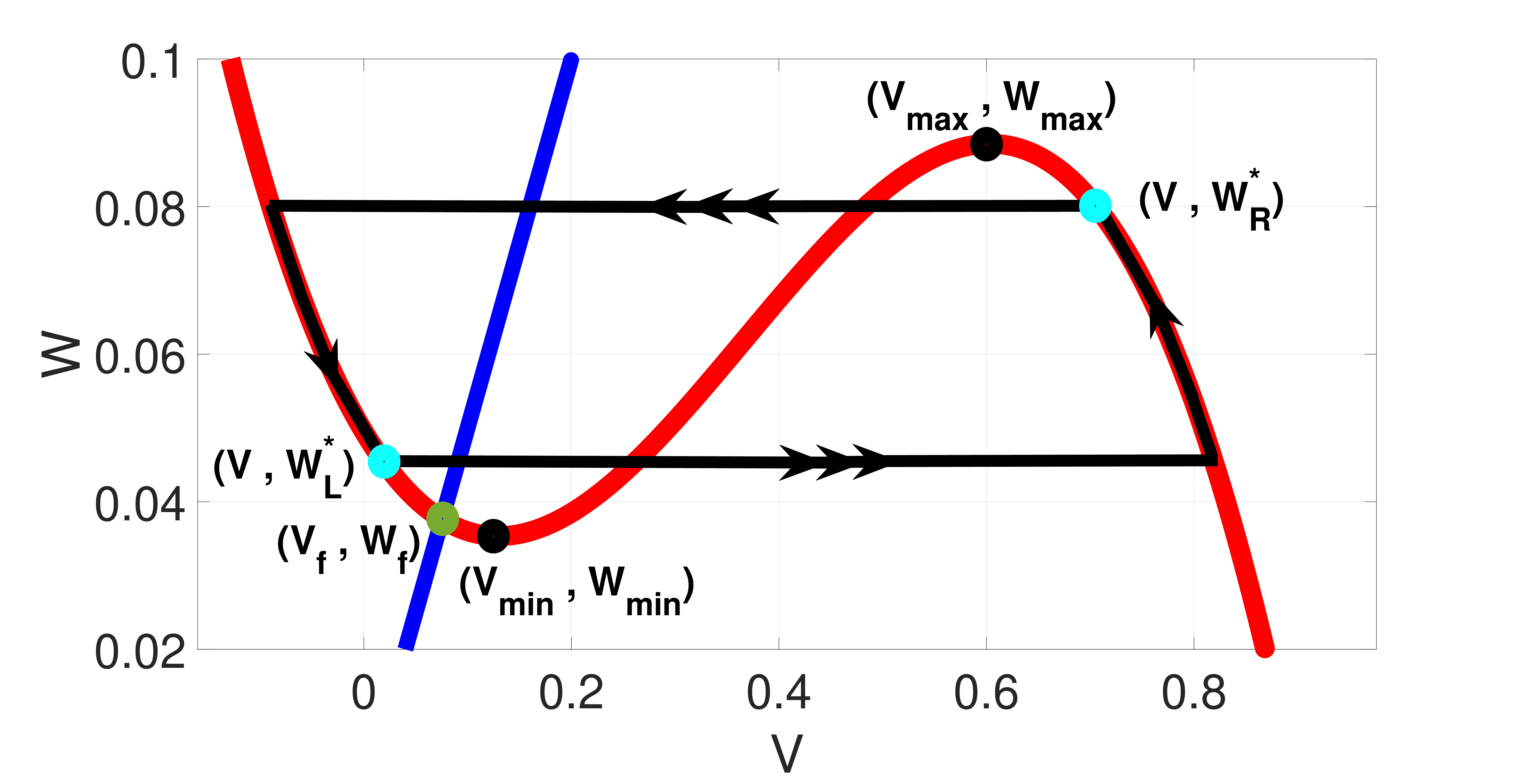}
    \caption{
    $W$-nullcline (blue line) and $V$-nullcline (red curve) of Eqs.~\eqref{eq:5} intersect at a unique fixed point $(V_f,W_f)$.
    Note that if $V_f<V_\min$, then $(V_f,W_f)$ is stable and in addition, if $W \in [W_\min, W_\max]$, then $W_L^*,W_f\in[W_\min,W_\max]$. The black loop represents a typical stochastic trajectory induced by SISR, where the horizontal parts with triple arrows indicate the fast escape at the points $W_{L}^{*}$ and $W_{R}^{*}$ from the left and right stable branches of the $V$-nullcline, respectively. The almost vertical parts of the trajectory, with single arrow, represent the slow motion of $W$ governed by Eqs.\eqref{eq:6c}. Note that $W_{L}^{*}>W_f$. $A=0.1$, $M=0.045$.}
\label{Fig:1}
 \end{figure}

Inserting $V=V^{*}_{L}(W)$ and $V=V^{*}_{R}(W)$ in the equation for $W$ in Eqs.~\eqref{eq:5} gives
\begin{eqnarray}\label{eq:6c}
%\begin{split}
    \left\{
    \begin{array}{lcl}
    \displaystyle{\frac{dW}{dt}} &=& \varepsilon \big[ bV^{*}_{L}(W) - cW \big],\\[3.0mm]
    \displaystyle{\frac{dW}{dt}} &=& \varepsilon \big[ bV^{*}_{R}(W) - cW \big] .
    \end{array}
    \right.
%\end{split}
\end{eqnarray}
The first (second) equation of Eqs.~\eqref{eq:6c} together with equation $V^{*}_{L}(W)$ ($V^{*}_{R}(W)$) governs the slow motion of $W$ down (up) the left (right) stable branch of the $V$-nullcline (see Fig. \ref{Fig:1}) to the leading order arising on the  $\mathcal{O}(\varepsilon^{-1})$ time scale when $\varepsilon\to0$.

Now, if we switch on the noise, i.e., $\eta_{_G}(t)\neq0$ with a small amplitude, $0<\sigma_n\ll1$, 
the first equation of Eqs.~\eqref{eq:6c} is not valid all the way down to the stable fixed point $(V_f,W_f)$ (in fact, for SISR to occur, the point $(V_f,W_f)$ should never be reached, otherwise, the trajectory would be trapped in the basin of attraction of the stable fixed point for a long time, thereby invoking a Poissonian spike train, leading to the non-occurrence of SISR) which is located on the left stable branch of the $V$-nullcline, i.e., $V_f<V_{min}$ (see Fig.~\ref{Fig:1}). 
But the first equation of Eqs.~\eqref{eq:6c} still governs the slow motion of $W$ until the well-defined point $W_{L}^{*}>W_f$ where a horizontal escape (invoked by noise) of a trajectory from the left stable branch of the $V$-nullcline occurs.

The same dynamics occur for the second equation of Eqs.\eqref{eq:6c} except that horizontal escape from the right stable  branch of the $V$-nullcline certainly occurs with or without noise. This is because the right (unlike the left) stable  branch of the $V$-nullcline has no fixed point to trap the trajectories and destroy the regularity of spikes. Thus, our analysis focuses only on the stochastic dynamics of the trajectories on the left stable branch.

To understand the escape mechanism of a trajectory from the left stable branch of the $V$-nullcline at the point $W_{L}^{*}$, we consider the limit $\varepsilon \to 0$, where the time scale separation between $V$ and $W$ becomes very large and  Eqs.~\eqref{eq:5} reduce to the 1D Langevin equation
\begin{equation}\label{eq:6}
\displaystyle{\frac{dV}{dt}}=-\frac{\partial U(V,W)}{\partial V} +  \eta_{_{G}}(t).
\end{equation}

In this limit, $W$ which comes from the solution of the first equation of Eqs.\eqref{eq:6c} is practically frozen and can be considered as a fixed parameter, its time variation providing only a $\mathcal{O}(\varepsilon)$ contribution to the dynamics governed by Eq.~\eqref{eq:6}. The function $U(V,W)$ in Eq.~\eqref{eq:6} is an effective double-well potential parametrically dependent on $M$:
\begin{eqnarray}\label{eq:7}\nonumber
    U(V,W)  &=& \frac{V^4}{4} - \frac{(1+A)}{3}V^3 +\frac{(3M+A)}{2}V^2 \\
            &-& [W - M(1+A)]V.
\end{eqnarray}

\begin{figure}
    \includegraphics[width=8.6cm,height=4.5cm]{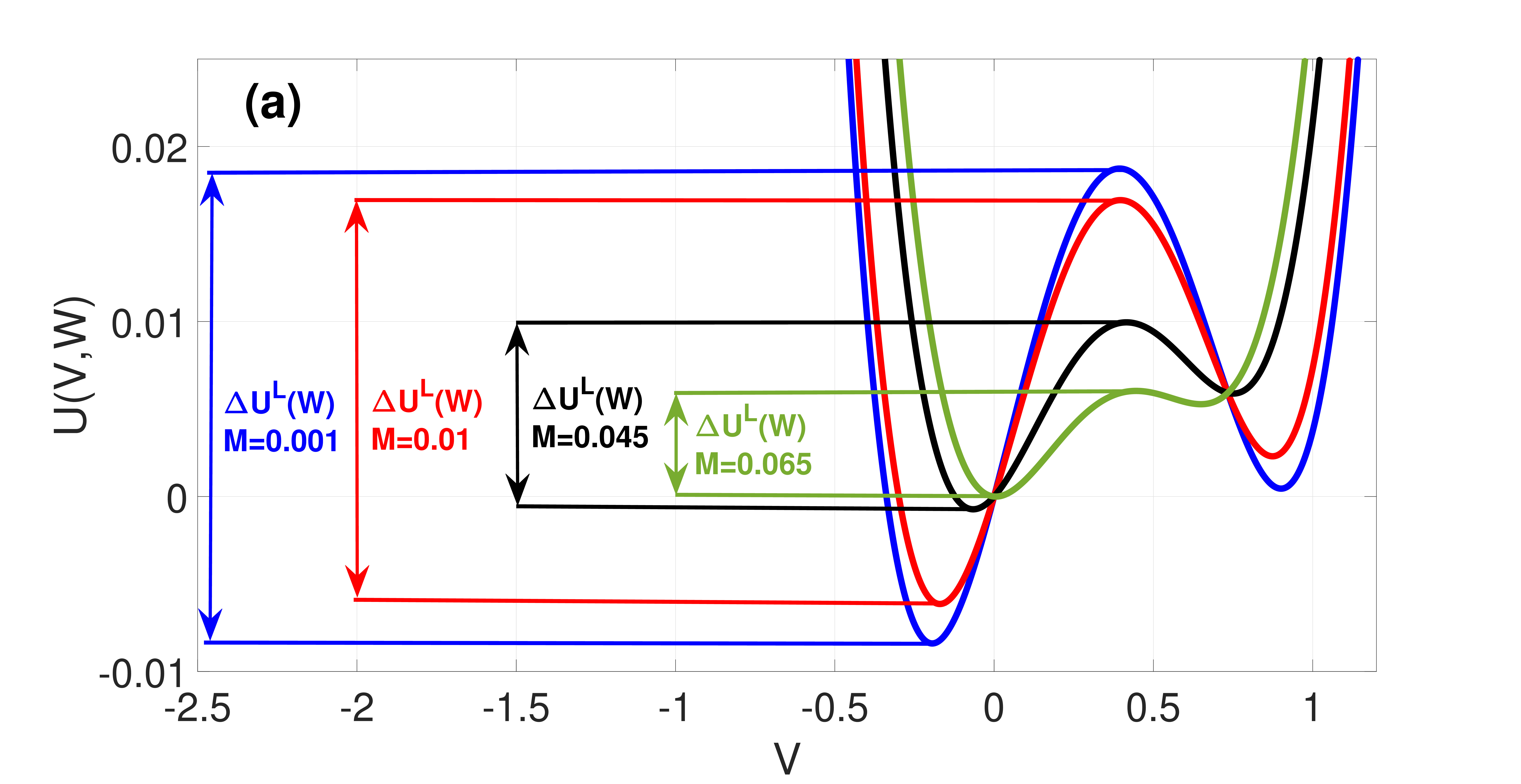}\\
    \includegraphics[width=8.6cm,height=4.5cm]{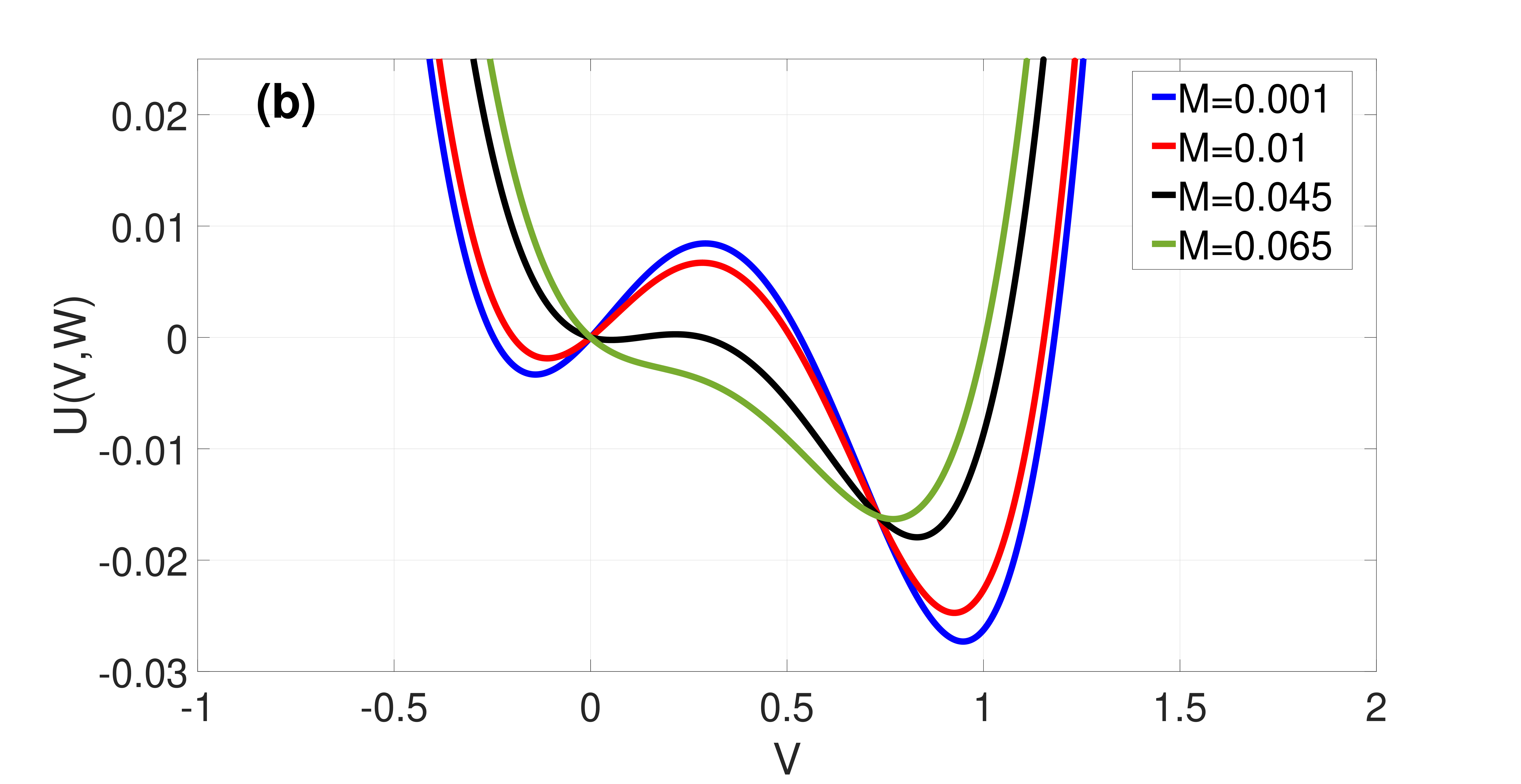}\\
    \includegraphics[width=8.6cm,height=4.5cm]{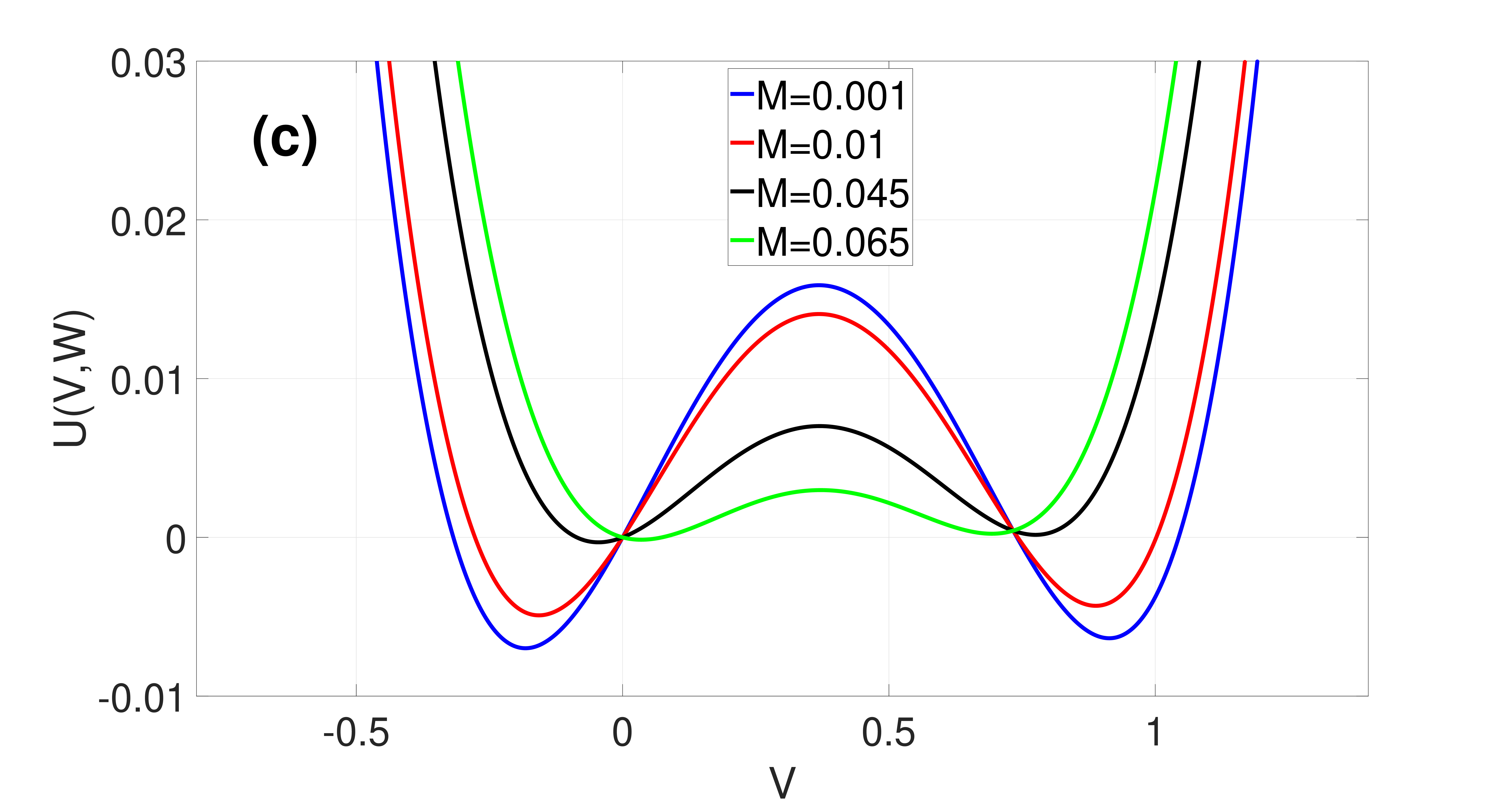}
    \caption{Landscape of $U(V,W)$ and energy barriers $\Delta U^{L,R}(W)$ for different values of $M$. 
    Panel (a): $U(V,W)$ is asymmetric ($\Delta U^{L}(W)>\Delta U^{R}(W)$) when $W(=0.07)>W_s$. Panel (b): $U(V,W)$ is asymmetric ($\Delta U^{L}(W)<\Delta U^{R}(W)$) when $W(=0.04)<W_s$. Panel \textbf{(c)}: $U(V,W)$ is symmetric ($\Delta U^{L}(W)=\Delta U^{R}(W)$) at $W_s=0.0621>W_f=0.0376$. $A=0.1$.}
    \label{Fig:2}
\end{figure}

\begin{figure}
\includegraphics[width=8.6cm,height=4.0cm]{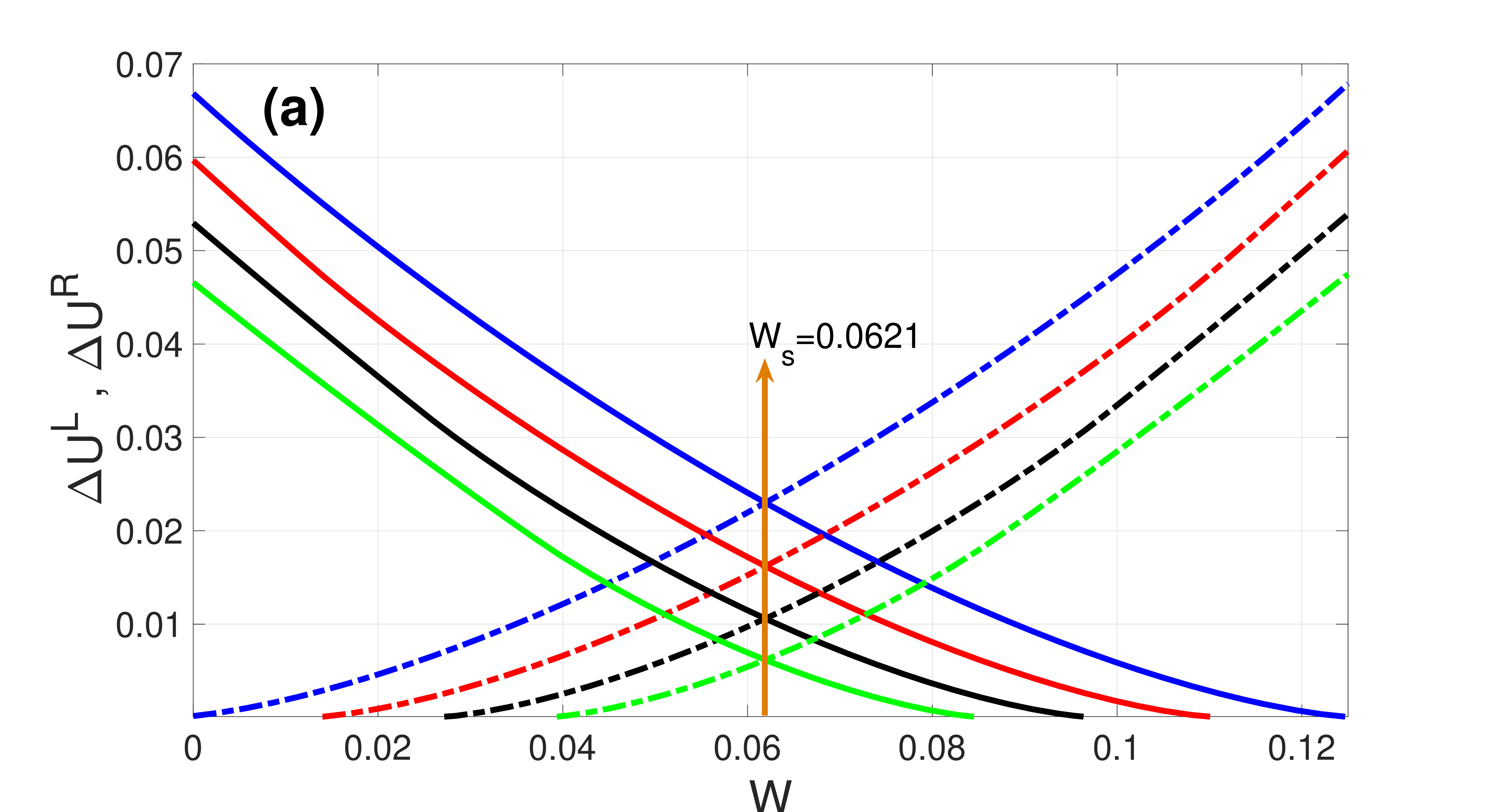}\\
\includegraphics[width=8.6cm,height=4.0cm]{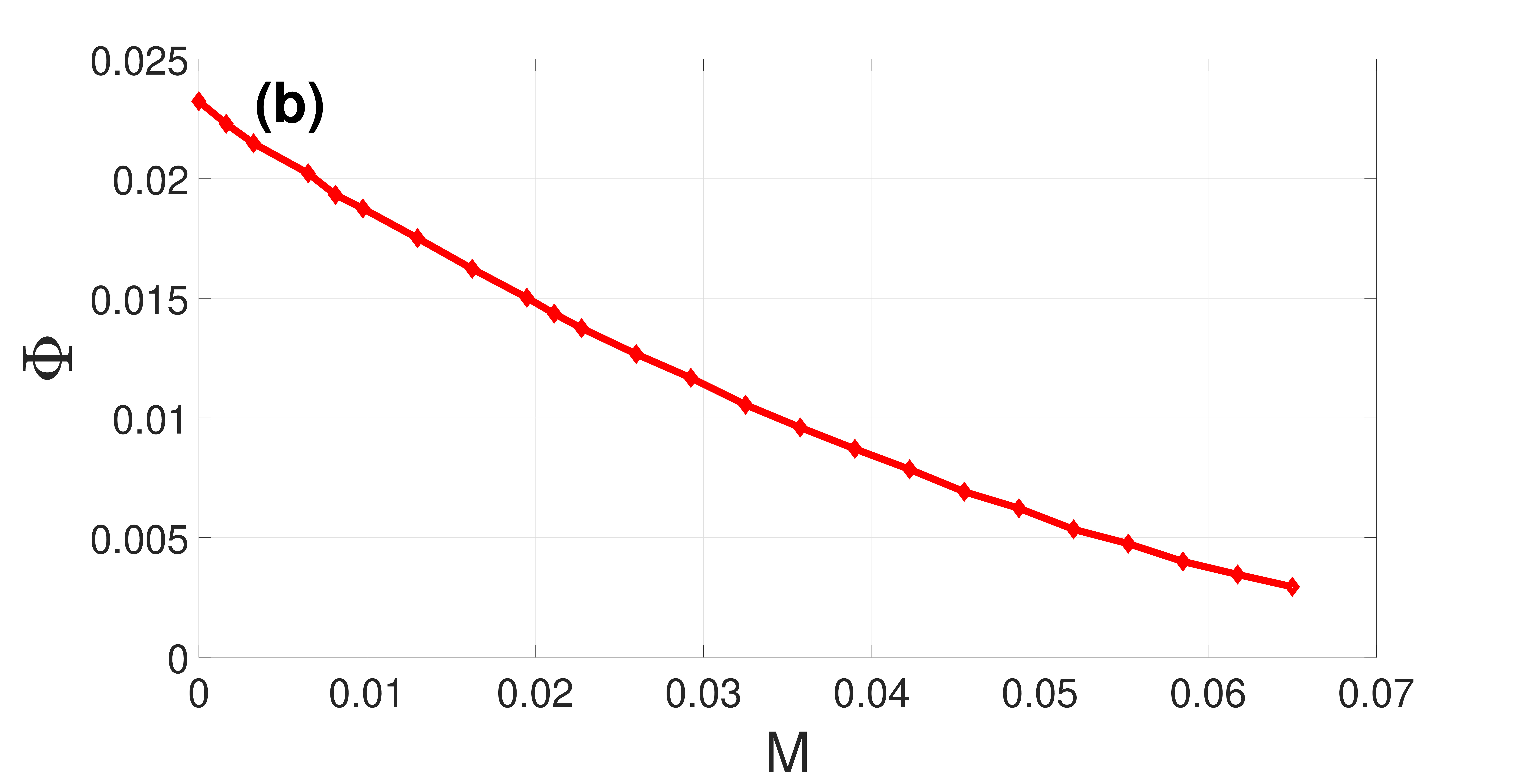}
\caption{Panel \textbf{(a)}: Variation of $\Delta U^{L}$ (dashed lines) and $\Delta U^{R}$ (solid lines) versus $W$ intersecting at $W=W_s=0.0621$ for values of $M=\{0.001,0.01,0.045,0.065\}$ shown in Fig.~\ref{Fig:2}. 
Panel \textbf{(b)}: Variation of $\Phi$ versus $M$. $A=0.1$.}
\label{Fig:3}
\end{figure}

Based on large deviations theory \cite{freidlin2001stable,freidlin2001stochastic} and Kramers’ law \cite{kramers1940brownian}, we write down for Eqs.~\eqref{eq:5} the generic conditions for the occurrence of SISR in slow-fast dynamical systems in the standard form \cite{yamakou2018weak,kuehn2015multiple} as follows \cite{muratov2005self,deville2007nontrivial,yamakou2018coherent}
\begin{eqnarray}\label{eq:8}
%\begin{split}
    \left\{\begin{array}{lcl}
        V_f < V_\min,\\[1.0mm]
        \displaystyle{\lim\limits_{(\sigma_{n},\varepsilon)\to(0,0)}\bigg[\frac{\sigma_{n}^2}{2}\ln(\varepsilon^{-1})\bigg]\in\Big(\Delta U^{L}(W_L^*), \Phi\Big)},\\[3.0mm]
        W_{L}^{*}>W_f,\\[1.0mm]
        \Delta U^{L}(W), \Delta U^{R}(W) \nearrow W\in[W_\min,W_\max].
    \end{array}\right.
%\end{split}
\end{eqnarray}
Here, ($V_\min,W_\min$) and ($V_\max,W_\max$) are, respectively, the minimum and maximum points of the $V$-nullcline, ($V_f$,$W_f$) is the unique (and stable) fixed point of Eqs.~\eqref{eq:5}, and $W_L^*$ is the value of $W$ that satisfies the first equation of Eqs.\eqref{eq:6c} and at which the trajectories escape almost surely from the left stable branch of the $V$-nullcine.
The left ($\Delta U^{L}(W)\ge0$) and right ($\Delta U^{R}(W)\ge0$) energy barriers of $U(V,W)$ are
\begin{eqnarray}\label{eq:12}
%\begin{split}
    \left\{\begin{array}{lcl}
\Delta U^{L}(W) = U\big(V^{*}_{S}(W),W\big) - U\big(V^{*}_{L}(W),W\big),\\[1.5mm]
\Delta U^{R}(W) = U\big(V^{*}_{S}(W),W\big) - U\big(V^{*}_{R}(W),W\big), 
    \end{array}\right.
%\end{split}
\end{eqnarray}
which are both non-negative and monotonic functions of $W$, see Fig.~\ref{Fig:3}(a).
% -------------------
 % -------------------
Figure~\ref{Fig:2} shows the landscape of $U(V,W)$ and how $\Delta U^{\{L,R\}}(W)$ varies with $M$. 
We note that the asymmetry of $U(V,W)$ is governed by $W$ and the double-well tends to disappear upon increasing $M$, resulting in a loss of the bistability required for SISR occurrence.
And $\Phi$ represents the intersection point of $\Delta U^{L}(W)$ and $\Delta U^{R}(W)$ at $W_s$, a point at which the two energy barriers are equal to each other. This happens when $U(V,W_s)$ is symmetric at $W_s>W_f$, i.e.,
\begin{equation}\label{eq:15}
 \Phi:=\Big\{\Delta U^{L}(W_s) : \Delta U^{L}(W_s) = \Delta U^R(W_s), W_s>W_f\Big\}.
\end{equation}
At the point $W_s$, the escape of a trajectory $V$ from the left stable branch and from the right stable branch of the $V$-nullcline are both equally less probable.

% -------------------
 % -------------------
 
In \eqref{eq:8}, the first condition ensures that the fixed point is unique and stable; the second condition ensures that a trajectory can escape (almost surely) from the left stable branch of the $V$-nullcline at the escape point $W=W_L^*$; the third condition ensures that the trajectory escapes before it reaches the stable fixed point, so that it does not get trapped into the basin of attraction of this fixed point for too long; and in the fourth condition, the monotonicity of $\Delta U^{L}(W)$ and $\Delta U^{R}(W)$ in the interval $[W_\min,W_\max]$ ensures that the escape points $W_L^*$ and $W_R^*$ on the left and right stable branches of the $V$-nullcline are unique, which would in turn ensure the periodicity of the trajectory leading to coherent spiking.

Since $W_f$ is the lowest attainable point of a trajectory on the left stable branch of the $V$-nullcline and the interval $\big(\Delta U^{L}(W_f), \Phi\big)$ in the second condition in \eqref{eq:8} is open, SISR deteriorates (i.e., the spiking becomes less coherent) and eventually disappears moving away from the center of the interval. Thus, for a given $\varepsilon \ll 1$, we use the boundaries of this interval to calculate the minimum ($\sigma_{n}^{\min}$) and maximum ($\sigma_{n}^{\max}$) noise intensity between which the highest degree of SISR can be achieved:
\begin{equation}\label{eq:14}
    \sigma_{n}^{\min} = 
    \displaystyle{\sqrt{\frac{2\Delta U^L(W_f)}         
                {\ln(\varepsilon^{-1})}}} \hspace{0.2cm}, \hspace{0.2cm}
    \sigma_{n}^{\max} = 
    \displaystyle{\sqrt{\frac{2\Phi}{\ln(\varepsilon^{-1})}}} .
\end{equation}
The quantities $\sigma_{n}^{\min}$ and $\sigma_{n}^{\max}$ have a dependence on the diversity parameter $M$ through $U(V,W)$ and $V^{*}_{L,S,R}(W)$. 
Thus, the length of the interval $(\sigma_{n}^{\min}$, $\sigma_{n}^{\max})$ can be controlled by $M$. 
It is worth noting that when $\sigma_n=0$, diversity alone cannot induce SISR. This is because, no single neuron in the network can spike as long as the excitability parameter (which is also the heterogeneity parameter) $a_i \sim \mathcal{N}(a_m,\,\sigma_d)$ lies in $(0,1+\sqrt{2})$, i.e., the excitable regime.

The occurrence of SISR  depends on whether the parameter values of the system, including $M$, satisfy the four conditions  \eqref{eq:8} in the double limit $(\sigma_{n},\varepsilon) \to (0,0)$. 
Hence, it suffices to study the variation of $\Phi$ versus $M$ to uncover the effect of diversity on the degree of SISR.
This is done in Fig.~\ref{Fig:3}, 
showing that $\Phi$ decreases upon increasing $M$. 
Thus, DIDC occurs when diversity in the network increases, leading to a deterioration and eventually destruction of the coherence of the spike train due to SISR, by shrinking the length of the interval $(\sigma_{n}^{\min}$, $\sigma_{n}^{\max})$ toward zero. 
% -------------------
% -------------------

%%%%%%%%%%%%%%%%%%%%%%%%%%%%%%%%%%%%%%%%%%%%%%%%%%%%%%%%%%%%%%%%%%%%%%%%%%%%%%%%%%%%%%%%%%%%%%%%%%%%
%\section{Numerical simulations}\label{Sec. IV}
%%%%%%%%%%%%%%%%%%%%%%%%%%%%%%%%%%%%%%%%%%%%%%%%%%%%%%%%%%%%%%%%%%%%%%%%%%%%%%%%%%%%%%%%%%%%%%%%%%%%
We corroborate the theoretical analysis via numerical simulations. 
We numerically integrate Eqs.~\eqref{eq:1} for $N=100$ neurons using the fourth-order Runge-Kutta algorithm for stochastic processes \cite{kasdin1995runge} and the Box-Muller algorithm \cite{knuth1973art}.
The integration time step is $dt = 0.01$ and the total simulation time is $T=1.5\times10^{6}$. 
For each realization, we choose for the $i$th neuron random initial conditions  $[v_i(0),w_i(0)]$, with uniform probability in the ranges $v_i(0)\in(-1, 1)$ and $w_i(0)\in(0.2, 1)$. 
After an initial transient time $T_0=2.5\times10^{5}$, we start recording the neuron spiking times $t_i^{\ell}$ ($\ell\in\mathbb{N}$ counts the spiking times).
Averages are taken over 15 realizations, which warrant appropriate statistical accuracy. 

We illustrate the effect of diversity, synaptic noise, and distance of the excitable network from the oscillatory regime, measured by $\sigma_d$, $\sigma_{n}$, and $a_{m}$, respectively, on the degree of coherence of the spikes induced by SISR. 
We use the coefficient of variation ($\mathrm{cv}$) given by the normalized standard deviation of the mean interspike interval (ISI) \cite{pikovsky1997coherence}.
For $N$ coupled neurons, $\mathrm{cv}$ is given by \cite{masoliver2017coherence}
\begin{equation}\label{eq:15a}
    \mathrm{cv} =
    \frac{\sqrt{\overline{\langle \mathrm{\tau}^2 \rangle} - \overline{\langle \mathrm{\tau} \rangle^2}}}
        {\overline{\langle \mathrm{\tau} \rangle}},
\end{equation}
where
$\overline{\langle \mathrm{\tau} \rangle} = N^{-1} \sum_{i=1}^N \langle \mathrm{\tau}_i \rangle$ and 
$\overline{\langle \mathrm{\tau}^2 \rangle} = N^{-1}\sum_{i=1}^N \langle \mathrm{\tau}_i^2 \rangle$, with
$\langle \mathrm{\tau}_i \rangle$ 
and 
$\langle \mathrm{\tau}_i^2 \rangle$ representing
the mean and mean squared ISI (over time),  $\mathrm{\tau}_i = t_i^{\ell+1}-t_i^{\ell}>0$, of neuron $i$.
 
\begin{figure}
    (a)\\\includegraphics[width=4.25cm,height=3.5cm]{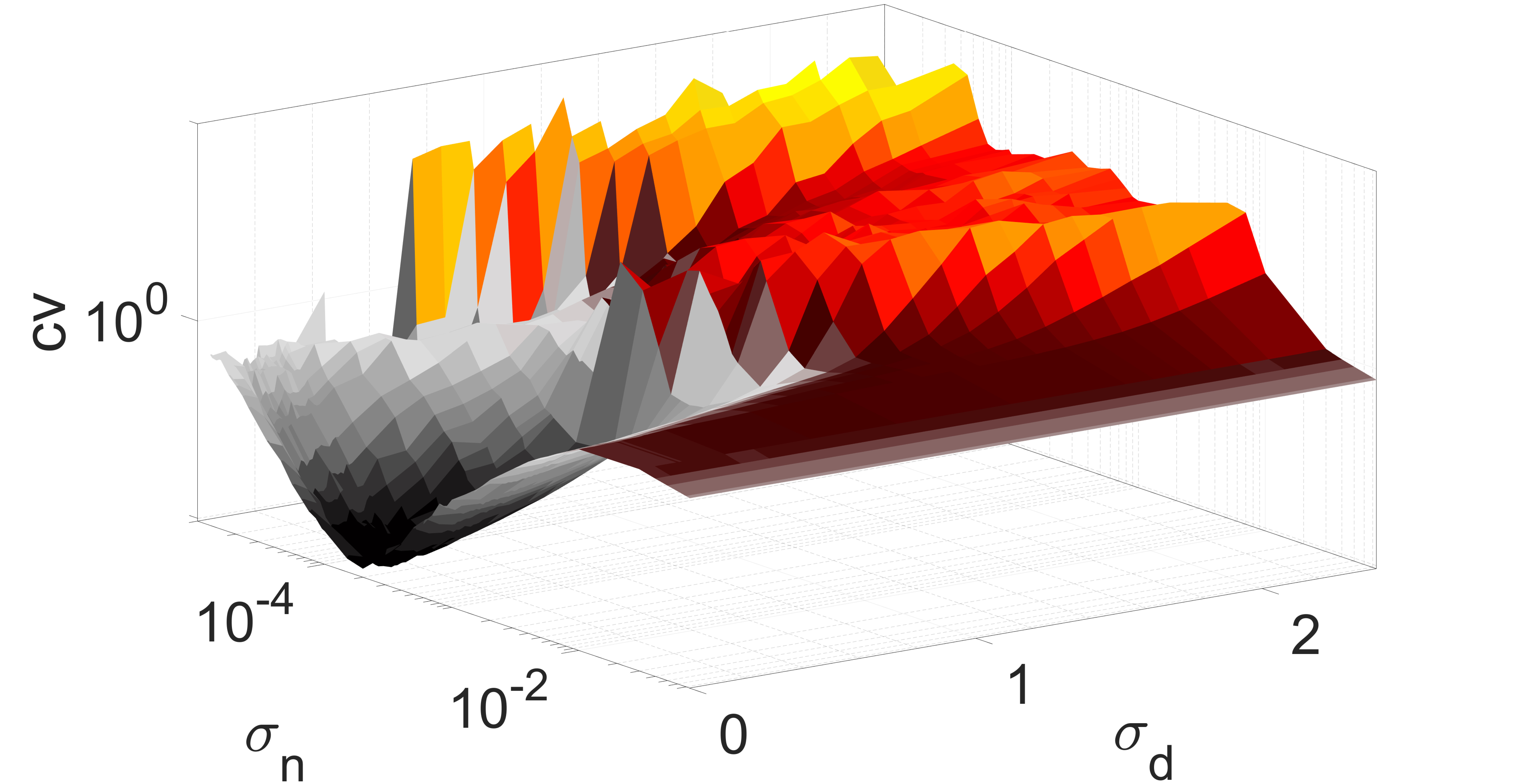}\includegraphics[width=4.25cm,height=3.5cm]{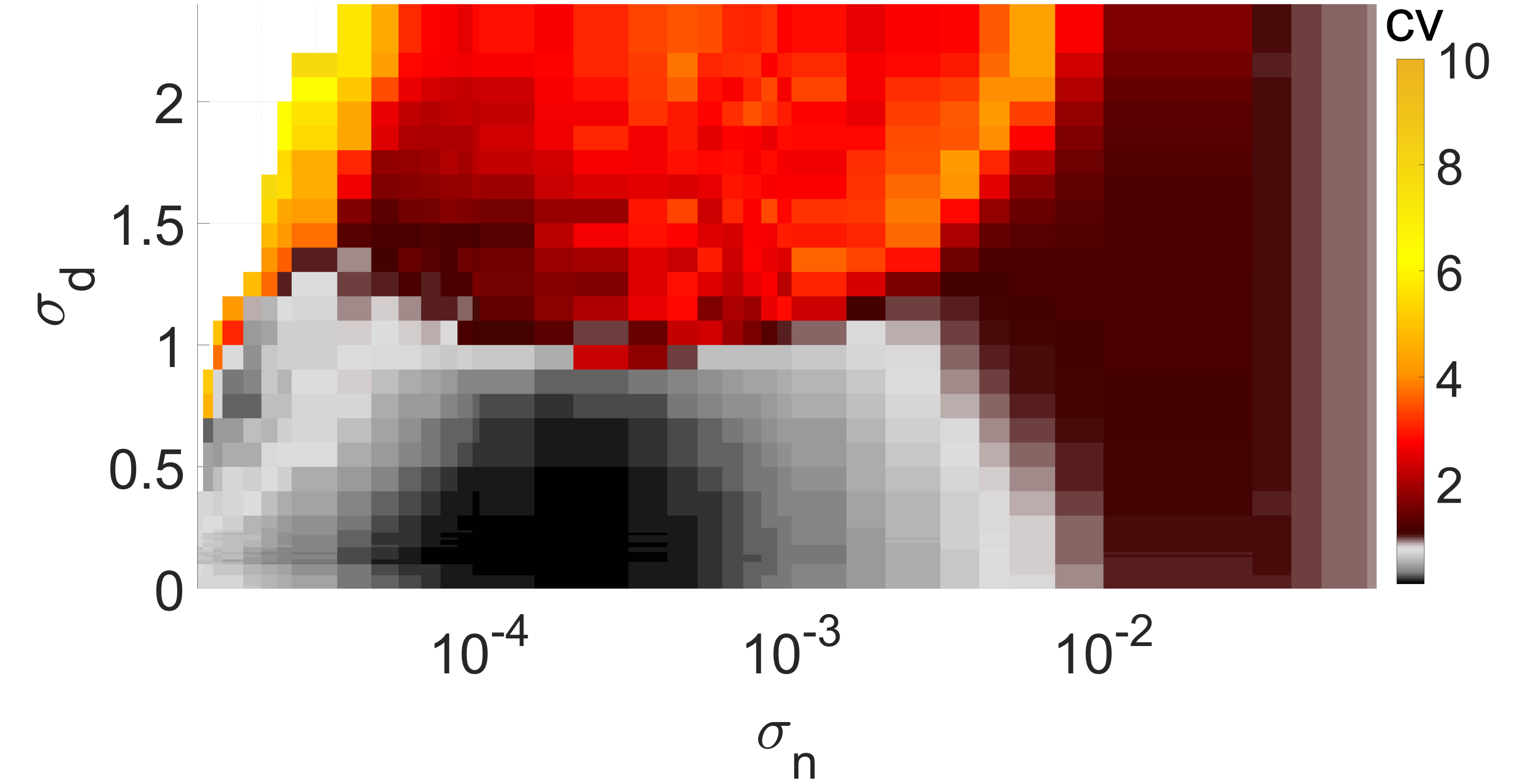}\\
      (b)\\\includegraphics[width=4.25cm,height=3.5cm]{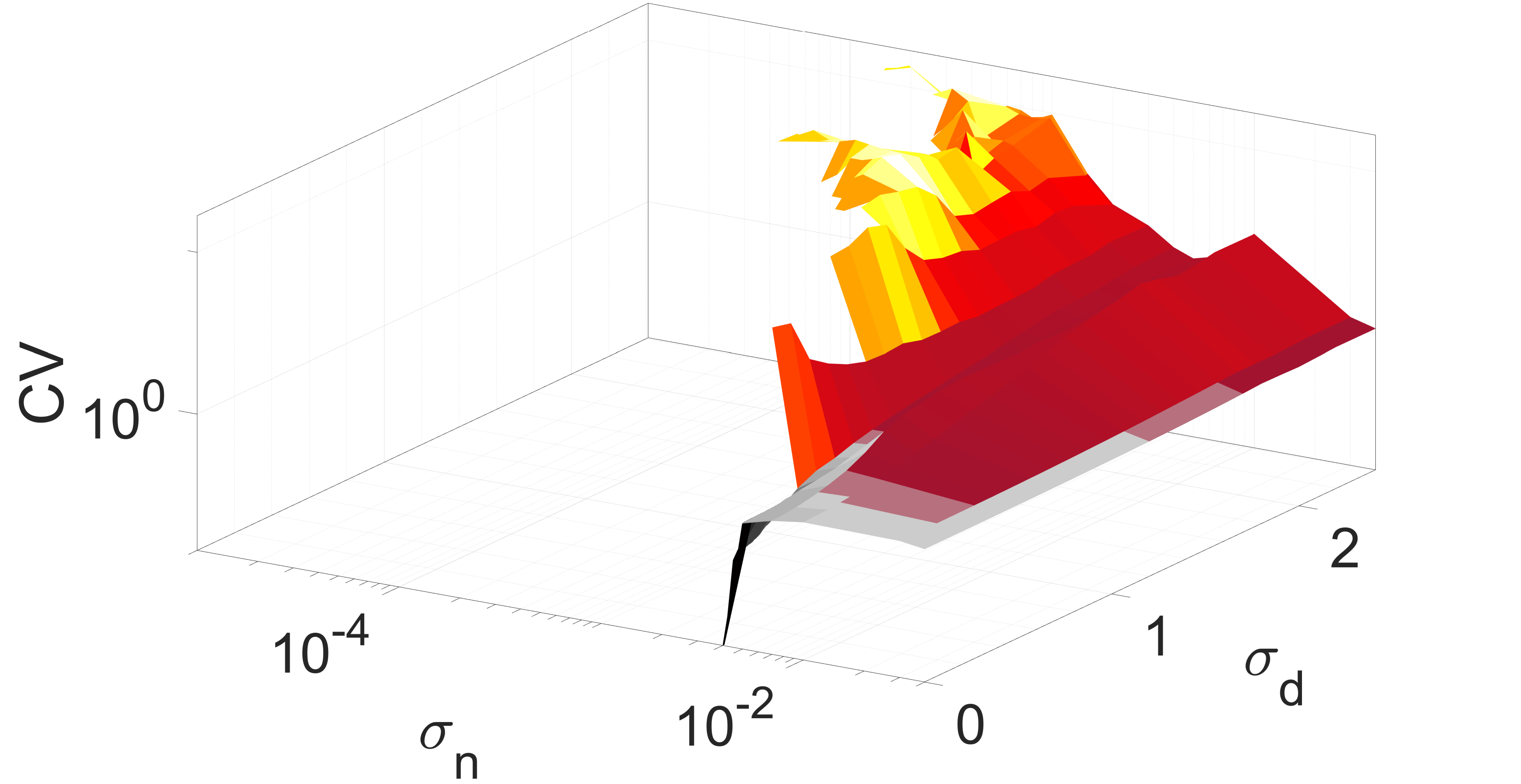}\includegraphics[width=4.25cm,height=3.5cm]{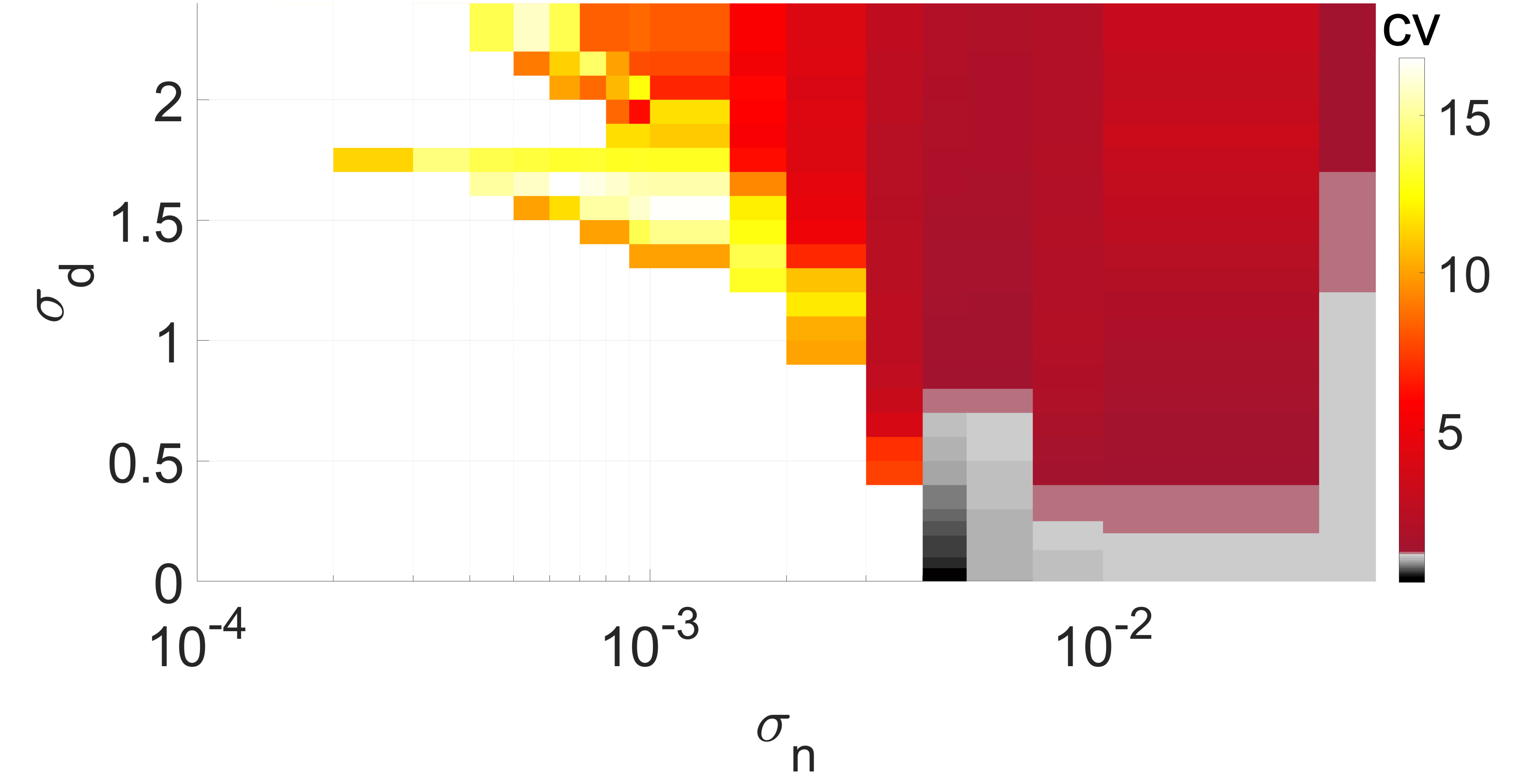}
    \caption{Panel (a): $\mathrm{cv}$ versus $\sigma_{n}$ and $\sigma_{d}$ in 3D with the 2D projection onto $(\sigma_n,\sigma_d)$-plane when  $a_m=0.05$. Panel (b): $\mathrm{cv}$ versus $\sigma_{n}$ and $\sigma_{d}$ in 3D with the 2D projection onto $(\sigma_n,\sigma_d)$-plane when  $a_m=1.2$. In both panels, the black and grey colors indicate values of $\mathrm{cv}<1$. Larger values of $\sigma_{d}$ inhibit SISR leading to larger $\mathrm{cv}$ values.}
\label{Fig:4}
\end{figure}

We determine the spike occurrence times from the instant the membrane potential variable $v_i$ crosses the threshold $v_{\mathrm{th}}=0.3$.  
The $\mathrm{cv}$ will be the higher the more variable the mean ISIs are. Thus, since Poisson spike train events are independent and all have a normalized standard deviation of unity (i.e., $\mathrm{cv}=1$), they can be used as reference for the average variability of spike trains of the network \cite{gabbiani1998principles}. 
When $\mathrm{cv}>1$, the average variability of spike trains of the network is more variable than a Poisson process. When $\mathrm{cv} < 1$, the average spiking activity of the network becomes more coherent, with $\mathrm{cv}=0$ corresponding to perfectly periodic spike trains. The degree of coherence is illustrated in Fig.~\ref{Fig:4}, which depicts $\mathrm{cv}$ against the synaptic noise $\sigma_{n}$ and diversity parameter $\sigma_{d}$ at two different values of $a_m$.

In Fig.~\ref{Fig:4}(a), the mean value $a_m=0.05$ is close to the lower bound of the interval $(0,1+\sqrt{2})$, i.e., close to the oscillatory regime. 
It can be observed that when $\sigma_{n} \in [10^{-4}, 10^{-3}]$ and $\sigma_{d}\in[0.0001,0.7)$, we have a low $\mathrm{cv}\in[0.107, 0.207]$, indicating a high degree of coherence due to SISR.
For $\sigma_{d} > 0.7$, the $\sigma_{n}$ interval in which $\mathrm{cv} <  0.207$ has shrunk to zero, i.e., $\mathrm{cv} \ge  0.276$ for all $\sigma_{n}$ values, indicating a significant deterioration and eventual destruction of the coherence as $\sigma_{d}$ increases.

In Fig.~\ref{Fig:4}(b), the mean of the diversity distribution is fixed at a higher value $a_m=1.2$.
In this case, the unique fixed point $(v_f,w_f)=(0,0)$ becomes even more stable 
than in Fig.~\ref{Fig:4}(a). 
Small diversities $\sigma_{d} \in [0.0001,0.3)$ and weak synaptic noise intensities $\sigma_{n} < 6\times10^{-3}$ are not strong enough to induce spiking; thus the network remains inactive and the value of $\mathrm{cv}$ is undefined. 

For $\sigma_{n} < 9\times10^{-4}$ and $\sigma_{d} > 2$, neurons respond differently to the synaptic noise due to the diverse strengths of the excitable regimes.
Due to the all-to-all coupling in the network, the large diversity boosts the weak synaptic noise, leading to the production of spikes. 
However, because the diversity is large, the conditions required for SISR are violated and the spikes produced are incoherent --- see in Fig.~\ref{Fig:4}(b) the magenta region bounded by $\sigma_{n} < 9\times10^{-4}$ and $\sigma_{d} \in [1.7,2.4]$, where $\mathrm{cv}>1.5$. 
At a relatively stronger synaptic noise intensity, i.e.,
$\sigma_{n} = 4\times10^{-3}$ and a very small diversity of $\sigma_{d}=0.001$, the degree of coherence due to SISR is best and $\mathrm{cv}=0.14$.
As $\sigma_{d}$ increases while the synaptic noise is fixed at $\sigma_{n} = 4\times10^{-3}$, the degree of SISR deteriorates and $\mathrm{cv} > 1$.

The results in Fig.~\ref{Fig:4} were obtained for a specific value of the time scale parameter ($\varepsilon=0.001$), which is a crucial parameter for SISR. Moreover, additional simulations performed for other values of $\varepsilon \ll 1$ and $K\in(0.025,1.0)$ (not shown) lead to qualitatively similar results.
% % -------------------
% -------------------

%%%%%%%%%%%%%%%%%%%%%%%%%%%%%%%%%%%%%%%%%%%%%%%%%%%%%%%%%%%%%%%%%%%%%%%%%%%%%%%%%%%%%%%%%%%%%%%%%%%%
%\section{conclusion}\label{Sec. V}
%%%%%%%%%%%%%%%%%%%%%%%%%%%%%%%%%%%%%%%%%%%%%%%%%%%%%%%%%%%%%%%%%%%%%%%%%%%%%%%%%%%%%%%%%%%%%%%%%%%%
In conclusion, we have provided evidence that there are complex network configurations and parameter regimes where diversity can only cause a deterioration of well-known resonance phenomena, such as SISR. 
This is predicted by our mean field analysis and confirmed by numerical simulations.

The decoherence effect appears as soon as there is a minimal degree of diversity in the system and rapidly grows up to a complete resonance muting as diversity increases.
The basic mechanism of this effect is that diversity causes a partial or complete disappearance of the energy barrier in the mean field double-well potential, responsible for the coherent spiking corresponding to SISR. 
The fact that in this system diversity cannot be optimized to enhance coherence, but can only disrupt it, is a nontrivial result. This is because the possibility to adjust diversity in order to amplify collective network behaviors has been previously demonstrated across a broad range of network types, configurations and conditions and is, therefore, a very general phenomenon~\cite{shibata1997heterogeneity,cartwright2000emergent,tessone2006diversity,toral2009diversity,chen2009threshold,wu2010diversity,wu2010effects,patriarca2012diversity,tessone2013diversity,grace2014pattern,patriarca2015constructive,liang2020diversity,kamal2015dynamic,scialla2021hubs,tessone2007theory,scialla2022interplay}.

We have illustrated the effect of DIDC in a prototypical excitable model network, which suggests that the effect may be common to other  physical, chemical, and biological systems. 
Based on our analysis and on experimental evidence that a neuron diversity loss can be associated to hyperkinetic disorders characterized by involuntary movements, we hypothesize that diversity may be used in biological systems not only to amplify weak signals, as suggested by previous literature, but also as an efficient control mechanism to prevent undesired resonances.
\vspace{2.0mm}

%%%%%%%%%%%%%%%%%%%%%%%%%%%%%%%%%%%%%%%%%%%%%%%%%%%%%%%%%%%%%%%%%%%%%%%%%%%%%%%%%%%%%%%%%%%%%%%%%%%%
%begin{acknowledgments}
M.E.Y acknowledges support from the Deutsche Forschungsgemeinschaft (DFG, German Research Foundation) -- Project No. 456989199. 
E.H., M.P, and S.S. acknowledge support from the Estonian Research Council through Grant PRG1059.
%\end{acknowledgments}
%%%%%%%%%%%%%%%%%%%%%%%%%%%%%%%%%%%%%%%%%%%%%%%%%%%%%%%%%%%%%%%%%%%%%%%%%%%%%%%%%%%%%%%%%%%%%%%%%%%%

%%%%%%%%%%%%%%%%%%%%%%%%%%%%%%%%%%%%%%%%%%%%%%%%%%%%%%%%%%%%%%%%%%%%%%%%%%%%%%%%%%%%%%%%%%%%%%%%%%%%
%\bibliography{references}% Produces the bibliography via BibTeX.
% %\bibliography{refs}{}
%\end{thebibliography}%
\providecommand{\noopsort}[1]{}\providecommand{\singleletter}[1]{#1}%
%

%%%%%%%%%%%%%%%%%%%%%%%%%%%%%%%%%%%%%%%%%%%%%%%%%%%%%%%%%%%%%%%%%%%%%%%%%%%%%%%%%%%%%%%%%%%%%%%%%%%%

\end{document}